\documentclass[aps,prl,twocolumn,tightenlines,superscriptaddress]{revtex4} 
\usepackage{epsfig}
\usepackage{graphicx}
\usepackage{psfrag}
\usepackage{amsmath,amssymb}
\usepackage{colordvi}
\usepackage{amsfonts}
\usepackage{enumerate}
\usepackage{slashed}

\begin{document}

\title{Comment on ``Does Gluons Carry Half of the Nucleon Momentum?" by X. S. Chen et. al. (PRL103, 062001 (2009))}
\author{Xiangdong Ji}
\affiliation{INPAC, Department of Physics, Shanghai Jiao Tong University, Shanghai, 200240, P. R. China}
\affiliation{Center for High-Energy Physics, Peking University, Beijing, 100080, P. R. China}
\affiliation{Maryland Center for Fundamental Physics, University of Maryland, College Park, Maryland 20742, USA}
\date{\today}
\vspace{0.5in}
\begin{abstract}
The authors claim to have found a ``proper", ``gauge-invariant" definition of a charged-particle's
momentum in gauge theory, which is more ``superior" than the textbook version. I show that 
their result arises from a misunderstanding of gauge symmetry by generalizing the 
Coulomb gauge result indiscriminately and is not physical. 
\end{abstract}

\maketitle

In a recent paper by Chen et al. \cite{Chen:2009mr}, the textbook definition of a charged-particle's 
momentum and angular momentum in gauge theory has been questioned. 
The authors claim they have found a ``proper" definition, and challenge
the well-known result in perturbative quantum chromodynamics (QCD) that the gluon 
carries one-half of the nucleon momentum in asymptotic limit. Here I argue 
that the textbook result stands, and the incorrect conclusion of the paper 
arises from a misunderstanding of gauge symmetry.  

In Ref. \cite{Chen:2009mr}, a ``sound" definition of a charged particle's momentum 
in a U(1) gauge field $A^\mu$ is purported to be (see Eq. (6)) 
\begin{equation} 
     \vec{P}_q = {\vec P} - q\vec{A}_{\rm pure}/c \ , 
\end{equation}
where ${\vec P}$ is the canonical momentum and $\vec{A}_{\rm pure}$ is ``a pure gauge term 
transforming in the same manner as does the full $A^\mu$" and always gives ``null field strength." 
This {\it magical} $\vec{A}_{\rm pure}$ allows a ``gauge-invariant" definition of $\vec{P}_q $ and 
``physical" $A^\mu_{\rm phys} = A^\mu - \vec{A}_{\rm pure}$. The authors claim that 
the quark's $\vec{P}_q$ shall be measurable in deep-inelastic scattering (DIS) and shall contribute 1/5 of the nucleon momentum. 

Electromagnetic theory is quite mature by now, and what is theoretically sound to define
and experimentally measurable is fairly well-known. The kinematic momentum of a
charge particle is 
\begin{equation}
    \vec{\pi} = {\vec P} - q\vec{A}/c \ , 
\end{equation}
with the full gauge field $A^\mu$ required! It is $\vec{\pi}$ which gives rise to the 
{\it kinetic energy} of the particle $E = \vec{\pi}^2/2m$, and it is $\vec{\pi}$ which generates
the {\it electric current}, $\vec{j} = (q/m)\vec{\pi}$. Feynman in his famous lectures provided
a beautiful example (Sec. 21-3) to demonstrate $\vec{\pi}$ is the momentum related to the velocity
of a charge particle measurable experimentally~\cite{feynman}. $\vec{\pi}$ is what measured in 
electron-nucleon DIS through QCD factorization. No physics is learned in separating  
$\vec{\pi}$ into ``gauge-invariant" combination 
\begin{equation}
 \vec{\pi} = \vec{P}_q - q\vec{A}_{\rm phys}/c \ , 
\end{equation}
there is no place for $\vec{P}_q$ in either experiment or theory, and ${\vec A}_{\rm phys}$ 
is never an observable in electromagnetism as $\vec{E}$ and $\vec{B}$ are!

A gauge field describes two dynamical degrees of freedom (d.o.f) of a 
massless spin-1 particle. A most economic description would have been
using a two-component field. However, to have Lorentz symmetry, 
one has to imbed the two d.o.f into a four-vector field, thereby introducing the gauge 
degrees of freedom. To ensure the gauge part do not contribute to physical observables,  
manifest gauge symmetry under $A^\mu \rightarrow A^\mu + \partial^\mu \chi$ is required. 
The gauge degrees of freedom seem to be a nuisance, it would be nice to get 
rid of them in actual calculations. However, this can only be done by 
first formulating a {\it Lorentz-invariant} theory and then imposing {\it gauge conditions}. 
The order of the procedure here is critically important and cannot be reversed: one cannot 
construct physical observables directly in term of ``physical" degrees of freedom after 
imposing the gauge conditions. Reversely-engineered gauge symmetry is not guaranteed physical 
because 1) observables generally do not have proper Lorentz transformation, 2) 
they generally are non-local, 3) they generally have no physical 
measurements. This, unfortunately, is exactly what Ref. \cite{Chen:2009mr} is advocating. 
Separating the gauge field into pure and physics parts is not Lorentz symmetric and cannot be done without 
destroying a local formulation of the theory. An alert reader would have found that this separation 
is practically the same as imposing the Coulomb gauge condition. What is then wrong with this gauge choice? Doesn't ${\vec  A}_\perp$ 
describe the physical degrees of freedom and other components of $A^\mu$ are the
pure gauge part? The answer is that there is nothing wrong with imposing the Coulomb gauge until 
one tries to {\it invent ``physical observables" with ${\vec A_\perp}$}. Colloquially, ${\vec A}_{\perp}$ is not 
``physical" enough such that anything made out of it is physical. 

DIS can never measure the ``gauge invariant" parton densities defined in \cite{Chen:2009mr} because they  
do not appear separately in factorization of the scattering cross section.  The main 
result of the paper (Eq. (8)) is obtained in Coulomb gauge and has no particular physical significance. 
The calculation is in fact 
incomplete because without the real gauge symmetry many other operators can mix
into $P_q$. This work was partially supported by the U. S. Department of Energy via grant DE-FG02-93ER-40762.

\end{document}